# MULTI-LOCUS INTERACTIONS AND THE BUILD-UP OF REPRODUCTIVE ISOLATION

Satokangas I[1], Martin SH[2], Helanterä H[3], Saramäki J[4], Kulmuni J[1,5]

1) Organismal & Evolutionary Biology Research Programme, University of Helsinki, Viikinkaari 1, P.O.Box 65, 00014 University of Helsinki, Finland
2) Institute of Evolutionary Biology, University of Edinburgh, Ashworth Laboratories, EH9 3FL, UK
3) Ecology and Genetics research unit, University of Oulu, P.O. Box 3000, 90014 University of Oulu, Finland
4) Department of Computer Science, P.O. Box 11000, FI-00076 AALTO, Aalto University, Espoo, Finland
5) Tvärminne Zoological Station, University of Helsinki, J. A. Palménin tie 260, 10900 Hanko, Finland

ORCID: IS, 0000-0002-2553-1487; SHM, 0000-0002-0747-7456; HH, 0000-0002-6468-5956; JS: 0000-0002-5904-4062; JK, 0000-0002-8852-0793

**Author for correspondence:** Ina Satokangas (ina.satokangas@helsinki.fi)

**Keywords:** multi-locus interactions, networks, speciation, reproductive isolation, epistasis, incompatibility

**Abstract**

All genes interact with other genes, and their additive effects and epistatic interactions affect an organism's phenotype and fitness. Recent theoretical and empirical work has advanced our understanding of the role of multi-locus interactions in speciation. However, relating different models to one another and to empirical observations is challenging. This review focuses on multi-locus interactions that lead to reproductive isolation (RI) through reduced hybrid fitness. We first review theoretical approaches and show how recent work incorporating a mechanistic understanding of multi-locus interactions recapitulates earlier models, but also makes novel predictions concerning the build-up of RI. These include high variance in the build-up rate of RI among taxa, the emergence of strong incompatibilities producing localised barriers to introgression, and an effect of population size on the build-up of RI. We then review recent experimental approaches to detect multi-locus interactions underlying RI using genomic data. We argue that future studies would benefit from overlapping methods like Ancestry Disequilibrium scans, genome scans of differentiation and analyses of hybrid gene expression. Finally, we highlight a need for further overlap between theoretical and empirical work, and approaches that predict what kind of patterns multi-locus interactions resulting in incompatibilities will leave in genome-wide polymorphism data.

**Introduction**

No gene works in isolation. Instead, genes interact with other genes and regulatory factors resulting in both additive effects and epistatic interactions (see Box 1 for definitions), which underlie phenotype and fitness. Consequently evolution, speciation and adaptation can be strongly influenced by interactions among multiple loci. In this review we are interested in the role of multi-locus interactions (see Box 1) in the build-up of reproductive isolation (RI). Much of the theoretical basis for understanding the evolution of RI is founded on epistasis and the concept of gene interactions. Bateson (1), Dobzhansky (2) and Muller (3) all suggested a model where two populations may become incompatible if they undergo substitutions at two interacting loci that are only 'tested' in hybrids (BDM incompatibilities or BDMIs, see Part 1). Gene interactions underlying RI are conventionally considered in the context of intrinsic RI, but interactions also play a role in the evolution of extrinsic isolation and divergence of 'ecological speciation genes'. Multi-locus interactions will tend to underlie most traits, including those under ecological selection, and strong selection on one locus could lead to co-evolution with other interacting loci. Multi-locus interactions play a role both in the beginning and later in the speciation process. However, as RI accumulates, the more loci are likely to be involved, and therefore the more important it is to consider the possible interactions between loci in contributing to strong RI.

Empirical evidence shows that the nature of multi-locus interactions, such as the structure of interaction networks (see Box 1), and co-evolution of interacting partners can have consequences for phenotype and fitness. For example, in the yeast genome, most genes interact with a fairly small number of others, but other genes are 'hubs' with very large numbers of interaction partners (4). Hub genes tend to have stronger fitness consequences when mutated in comparison to loci with fewer interactions (5–7) potentially leading to different consequences for RI. The structure of multi-locus interactions also influences where epistasis, and thus genetic incompatibilities, are likely to arise. Incompatibilities might be likely to involve loci that function in the same biological process or protein complex, as suggested by enrichment of negative fitness epistasis in a yeast mutation study (8).

Our growing empirical understanding of gene regulation and protein interactions has inspired conceptual expansions to the initial BDMI model. At the same time, genome-wide empirical studies are being utilised to understand the role of multi-locus interactions in RI, as discussed in Part 2. In this paper we bring together both theoretical and empirical approaches to understand how multi-locus interactions could drive the build-up of RI. Our review has three parts. First, we discuss different theoretical approaches that have been used to explore the possible roles of multi-locus interactions in speciation with an aim to link the different approaches at a conceptual level. We discuss their key findings and insights, and evaluate where additional modelling could further extend our knowledge. In the second part we turn to empirical studies and approaches that can shed light on the role of multi-locus interactions in the build-up of RI, highlighting the challenges with current methods. In the last part we conclude by identifying fruitful avenues for the future, particularly bringing together both theoretical and empirical work in this field and using genome-scale simulations as a tool to bridge between theory and observation.

**Box 1. Definitions.**

---

By **multi-locus interactions** we mean physical or statistical pairwise or higher order interactions among two or more genomic loci. We are especially interested in scenarios that involve interactions between more than two loci. By **loci** we mean either genes or regulatory sequences. Physical, direct interactions among the loci can take the form of protein-protein interactions (**PPI**), or regulation through interactions between DNA, RNA or protein molecules (9).

Our focus is on interactions that have an epistatic effect on fitness. By **epistatic effect on fitness** we mean that the fitness effect of an allele is dependent on alleles at other loci (or in some cases sites within a locus interact epistatically, e.g. (10,11)). For epistatically interacting alleles, their combined fitness effect deviates from linearity or additivity. Epistasis on the level of fitness can be caused either by epistatic effects on phenotypic traits, or by additive effects that themselves have nonlinear effects on fitness. Epistatic fitness effect could arise due to physical interaction among loci. However, physical interactions among the loci do not necessarily translate into epistatic effects for fitness and likewise, epistatic effects for fitness can arise between loci that do not interact at physical level. For instance, alleles at two genes that contribute to the same trait via different pathways, or act as upstream regulators of the same pathway, can have epistatic effects on fitness without directly interacting with each other. To date we know very little about how interactions between loci at the molecular level (e.g. regulatory or protein-protein interactions) translate into epistasis that is relevant for specific traits or individual fitness.

Interactions between loci can be represented as a network (see regulatory network in Figure 1). **A network** represents the configuration of direct pairwise interactions between the elements, so that network nodes are the interacting elements and links correspond to their interactions. Although only direct interactions are shown as links in the network, it also determines possible indirect interaction pathways: nodes that are connected via some path through the network can in principle influence one another. In most networks, such paths can be traced between almost all nodes. The paths are typically rather short, with only a few intermediate nodes (12,13), which can be thought to facilitate indirect interactions. There are also many other network features that play important roles for dynamics taking place on networks. One is the existence of highly connected **hubs** (a node that has many more connections than other nodes) and related **broad or power-law connectivity distributions**, reflecting heterogeneity in the number of interaction partners between nodes (see, e.g., (14)), which can have strong effects on processes mediated by the network (see, e.g., (15)). Further, similarly to most other real-world networks, biological networks are usually sparse (see, e.g., (13)). Note that the general term 'network' does not, however, necessarily imply that the above-described features are present.

---

**PART 1: Theoretical models: How could multi-locus interactions influence speciation?**

Models of post-zygotic isolation that incorporate epistatic interactions assume that mutations that have positive or neutral fitness effects in the parent populations can cause reduced fitness when combined in hybrids. Alleles from divergent populations may also be beneficial when combined in hybrids, but in this article we focus on interactions with reduced fitness. This outcome can be modelled in several different ways. Many models focus on the evolutionary processes and spatial setting by which reduced hybrid fitness can come about, but fewer studies modelling multi-locus interactions explore the population genetic outcomes under ongoing hybridisation. One of the challenges in speciation research is therefore to determine how the different models relate to each other and to natural populations studied by empiricists. In this section we discuss insights from models of speciation accounting for multi-locus interactions without functional information on interactions, as well as those that incorporate our growing mechanistic understanding of multi-locus interactions (i.e. PPI and regulatory networks). We highlight their main similarities and differences as well as conclusions about the emergence and maintenance of RI. We do not address the extensive modelling in the context of ecological speciation, where divergent selection on multiple loci that are each independently selected can lead to RI despite the homogenising effect of gene flow (16–18). In the models accounting for multi-locus interactions that we address here, speciation occurs as a side-effect of substitutions that may be fixed through drift or selection, depending on the model. Whether or not gene flow is present also varies among models (as described below), but there is no assumption of antagonism between local adaptation and gene flow, which is a common feature of ecological speciation models.

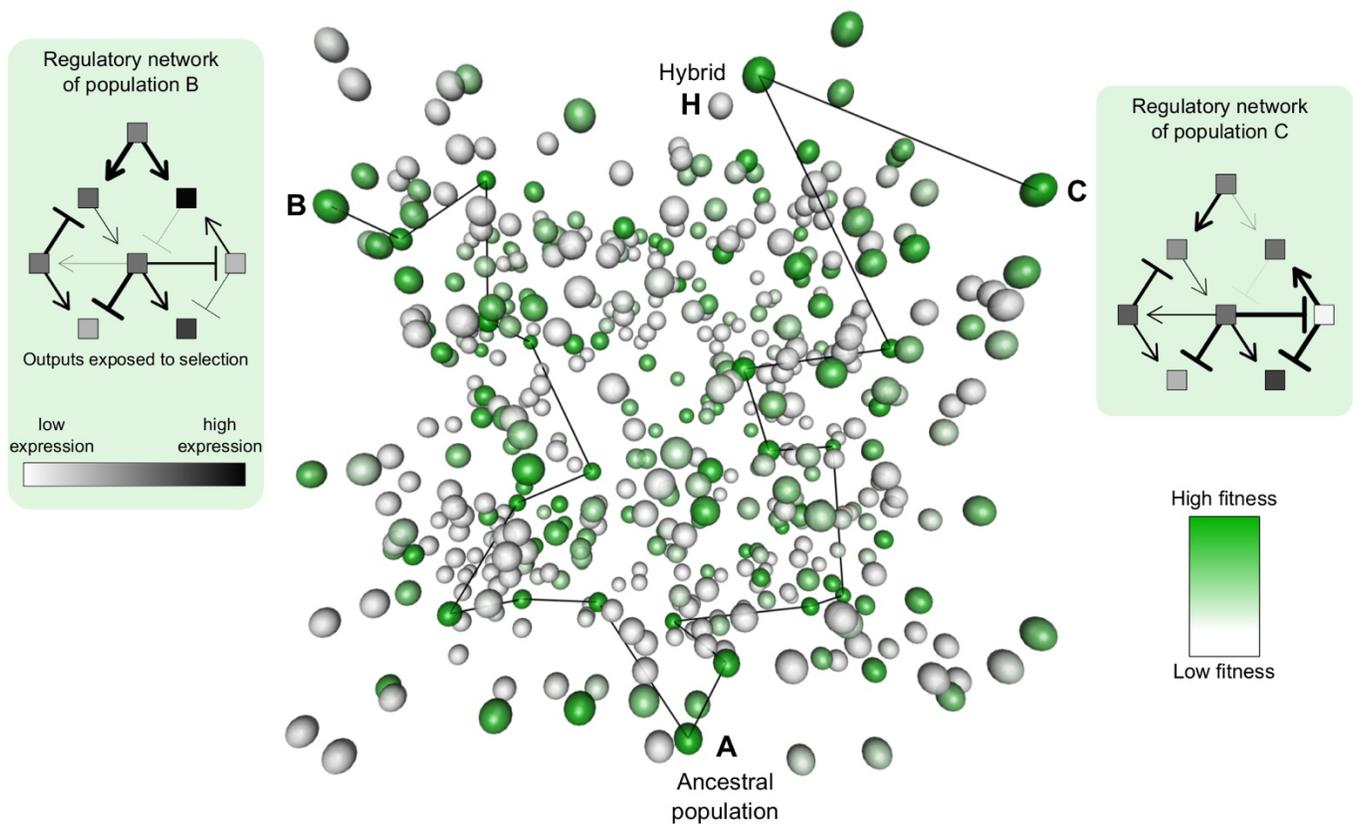

**Figure 1. Networks and fitness landscapes.** Conceptual representation of a high-dimensional fitness landscape with an example of regulatory network evolution leading to RI. Each sphere in the high-dimensional genotype cloud (central panel) represents a possible genotype combination for a multi-locus interaction network (see regulatory network in the side panel). However, not all possible genotype combinations are shown as in reality, the number of possible combinations will be extremely vast. The number of combinations arises because each gene can be mutated in a vast number of different ways, each with different effects on how they interact with other genes in the network. Colours indicate that fitnesses vary among genotype combinations. While fitness landscapes are usually depicted in two or three dimensions, in reality they represent much higher-dimensional space, in which each dimension corresponds to the range of possible genotypes at one locus. This high-dimensionality creates more opportunities for populations to traverse genotype space without crossing regions of low fitness ((19), and see the main text). Our depiction is intended to represent an arbitrary number of dimensions, although only three dimensions are plotted. Black lines indicate two possible paths along which the network could evolve while retaining high fitness from an ancestral population 'A' to daughter populations 'B' and 'C'. A black line between two spheres represents a single mutational step, so that spheres distant in the three plotted dimensions can be adjacent in a dimension that is not plotted. A hypothetical gene regulatory network is indicated for populations B and C. In the network, all genes can alter the expression of certain other genes through up (pointed arrows) or down (flat arrows) regulation. Fitness is determined by the expression levels of two output genes. A hypothetical recombinant hybrid individual 'H' is indicated to show that combining elements of the networks of populations B and C can result in reduced hybrid fitness.

*1.1 Epistasis underlying RI can be modelled without mechanistic knowledge of multi-locus interactions*

Probably the best-known class of models considers the direct impact of inter-locus interactions on hybrid fitness. In Bateson-Dobzhansky-Muller Incompatibilities (BDMIs) (1–3,20), substitutions that occur independently (either in two separate populations, or consecutively in one of the populations (21), through drift or selection) are never tested together, due to allopatry, until being combined in a hybrid, whereupon their interaction can lead to reduced fitness (i.e. epistasis). BDMI models have allowed predictions about how RI builds up over time. Assuming a fixed probability of a new substitution in population 1 leading to an incompatibility with an allele from population 2, incompatibilities should emerge at a rate proportional to the square of the number of differences between diverging populations. This leads to a faster than linear increase in the number of incompatibilities with genetic differences, the so-called 'snowball' effect (22,23). However, as these approaches tend to only consider the number of incompatibilities, they do not allow predictions about the fitness of recombinant hybrids (i.e. F2s or backcrosses, but see (24)). Therefore, they are limited in their ability to predict the strength of a barrier to gene flow in the face of ongoing hybridisation (25).

Fitness reduction in hybrids caused by new combinations of alleles was also modelled by Wright (26,27) using the idea of a fitness landscape, a multidimensional space in which each dimension represents the range of possible genotypes at a given locus. Each point in the landscape therefore represents a combination of alleles at multiple loci with a certain fitness. For a polygenic trait, there will inevitably be multiple combinations of alleles that combine to produce a phenotype of optimal fitness in a given environment, even when the alleles contribute to the phenotype additively. In other words, the landscape will have multiple 'peaks' with similar high fitness (27,28). These peaks can alternatively be represented as points of high fitness in a genotype space, as in Figure 1 (see also (29)). RI can arise between two populations that are both under stabilising selection for the same optimal phenotype, but which shift to different genotypic peaks of high fitness. This can occur through genetic drift followed by compensatory evolution at other loci (so called nearly neutral, or 'quasi-neutral' divergence) (27,28). Hybrids that carry a previously untested combination of genotypes may thus fall into a region of the landscape with lower fitness. If all loci affect a trait additively, F1 hybrids between parents with different high-fitness genotypes should also be fit, but F2s and other recombinant hybrids can be unfit due to segregation of distinct alleles from each parent that sum to an unfit phenotype when brought together (i.e. 'segregation variance' (30): a combination of additively acting alleles results in an epistatic effect on fitness). However, the emergence of RI under this model is slow (28) and at most linear with time (30). A more recent extension shows how adding explicit epistatic interactions in the phenotypic effects of pairs of loci can lead to faster than linear emergence of RI, with reduced F1 fitness as well as increased segregation variance in recombinant hybrids (31).

The visual representation of a fitness landscape with multiple peaks of similarly high fitness separated by deep 'valleys' has been criticised for failing to represent the true nature of a fitness landscape under high dimensionality. As the dimensionality of the landscape increases, i.e. when more loci contribute to fitness (as is the case for many traits), so does the probability that high fitness peaks are reachable through a series of small mutational steps without passing through regions of reduced fitness (19) (see Figure 1 for an example of a high-dimensional fitness landscape). In other words, peaks of high fitness are likely to be connected by neutral ridges in the higher dimensions. Gavrilets and Gravner (32) envisaged a 'holey landscape' represented as a flat plane interspersed with holes, thus implying that there are very many combinations of genotypes that have equally high fitness, but these are interspersed by combinations that have low fitness (32). As in the original BDMI models (1–3), the evolution of RI is not impeded by a requirement that parental populations traverse regions of low fitness. Increasing the dimensionality of the genotype space will make divergence by drift easier, so that in a multidimensional holey landscape, only a few substitutions may be necessary to result in RI and speciation (32).

A final approach for modelling the fitness consequences of hybridisation without the need to consider explicit mechanistic knowledge on multi-locus interactions is Fisher's Geometric Model (25,33,34). This approach defines the fitness landscape in terms of phenotype space rather than genotype space. Substitutions are represented as steps through phenotype space, either toward or away from the fitness optimum, potentially in more than one dimension due to pleiotropy. Since each substitution can happen at a different locus, this model allows the assumption of an infinitely large number of loci affecting fitness (25). The phenotype and fitness of any hybrid or backcross can be computed by combining phenotypic changes experienced by each of the parental populations. A key factor that affects the predictions of FGM is the shape of the fitness landscape. For example, Barton (25) assumed a quadratic decline in fitness away from the optimum, and found that strong incompatibilities are unlikely to arise by stabilising selection acting on the phenotype with drift changing the underlying genotype. However, Fraïsse et al. (34) showed that using a different shape of the fitness landscape with a plateau of high fitness allows the accumulation of larger effect substitutions that lead to stronger fitness decreases in recombinant hybrids. By adjusting the shape of the fitness landscape, FGM can account for many empirical patterns in speciation studies, including some that are not well explained by other models (34,35). One of these is that FGM predicts that severe loss of fitness occurs only if multiple factors are introgressed together, especially when recipient genotypes are well adapted (34). In contrast to Orr's (22) treatment of BDMIs, FGM does not predict the snowball effect, although it can generate an 'apparent snowball effect' when the introgressed regions contain a large number of divergent sites. Finally, FGM may be applied to answer questions in speciation that do not relate to incompatibilities per se, like whether divergence arose by drift or selection (36).

*1.2 Models that incorporate a mechanistic understanding of multi-locus interactions*

One possible criticism that applies to all of the models described above is that it is often difficult to make the connection from the theoretical model to the biological context in which loci interact to produce a phenotype (i.e. a mechanistic genotype-phenotype map). Below we discuss several recent efforts to model the emergence of incompatibilities while considering the nature of protein-protein or gene regulatory interactions and how they produce phenotypes. Such models can reveal whether adding functional and more mechanistic details of multi-locus interactions allows for additional insights.

*1.2.1 Incompatibilities can accumulate by drift in redundant gene regulatory networks*

A growing understanding of gene regulation has inspired mechanistic models that test whether evolution of regulatory networks can lead to RI (10,37–40). These models allow complex non-additive gene effects and are conceptually similar to fitness landscape models (27,28,31) in that fitness is determined by the combined effect of all loci contributing to a trait or collection of traits that are exposed to selection (i.e. the level or spatial distribution of expression of one or more genes). Because gene regulation tends to be somewhat redundant, populations can evolve and accumulate incompatibilities under directional selection towards the same optimum, by taking different mutational steps (38), or under stabilising selection, through drift and compensatory evolution (10,37,39). In the models of developmental and gene regulatory pathways, this process by which the underlying genetic basis of the trait can change even if the output phenotype remains the same has been termed 'system drift' ((41), see also (42)), but it is conceptually the same as drift under stabilizing selection in fitness landscape models (Section 1.1) (25). The key characteristic of regulatory networks that allow for this system drift is their redundancy, the fact that multiple genotypic states can lead to the same fit outcome. Schiffman and Ralph (40) show analytically that nearly all gene interaction networks are likely to share this property. While the role of system drift in regulatory networks has not been demonstrated in the context of RI, experimental studies have demonstrated the existence of networks of phenotypically stable genotypes within which a population may move (e.g. (43)), and rewiring of genetic pathways and changes in gene function underlying unchanged phenotypes (44).

*1.2.2 The number of connections and their distribution affect the rate and variance of accumulation of incompatibilities, and which genes are likely to be involved*

Orr's extension to the BDMI model (22) assumed a fixed probability of an incompatibility with each substitution in the diverging populations, meaning all loci could potentially interact with all others. However, this is probably unrealistic as not all genes within an organism interact with each other (but see (45)). Livingstone et al. (46) considered a network model based on the observed PPI network in yeast, with a broad connectivity distribution (i.e. power-law-like network topology, see Box 1), in other words some genes were hubs with many interactions and other genes were less interactive (see regulatory network in Figure 1). This leads to a slowing in the quadratic growth of incompatibilities by a factor that equals the fraction of node pairs that are connected in the network, *i.e.* the density of interactions, with lower density leading to slower increase in incompatibilities. Similarly, simulations (47) show that areas of the network with high densities of interactions are predicted to accumulate incompatibilities at faster rates in comparison to areas where interactions are sparse. However, this process may be opposed by slower rates of substitutions at the more highly connected nodes due to pleiotropic constraint (5), as has been repeatedly observed in both gene co-expression networks and PPI networks (see e.g. (7,48–50). Thus the rate of accumulation of incompatibilities may be dramatically slowed compared to the snowball model (51)). Another important consideration is the variance in the rate of accumulation of incompatibilities. Orr and Turelli (23) investigated the effect of stochasticity in substitution rate and effect size on fitness, but to date studies have not considered additional variance in the build-up of RI introduced by the power-law-like network topology. This type of network topology could increase the variance in the rate of accumulation of RI, compared to networks of a similar size, but with more evenly distributed connectedness. This means that species pairs could differ strongly in the rate at which RI builds up, despite similar network topologies. Thus, the structure of interaction networks alone could partly help explain why the strength of RI can differ dramatically between species pairs with similar genetic distance (52), without invoking any additional differences between the species pairs.

*1.2.3 Co-evolution within a network and within-population positive epistasis can create strong localised barriers that are persistent in the face of gene flow*

While early models of BDMIs (1–3) consider only the deleterious consequences of epistasis, several recent studies have explored the consequences of beneficial interactions, which could arise among loci that participate within the same physical interaction network (53,54). Co-evolution between interacting loci (55) could occur through multiple successive steps. The breaking of the resulting within-population 'positive epistasis' (53) in hybrids can lead to effective species barriers. This has been shown in population genetic studies that examine the fate of incompatibilities in the face of gene flow - a consideration that was often ignored in the past. Unlike BDMIs in their original formulation (where fit ancestral combination can be recovered by recombination), incompatibilities that result from multi-step co-evolution in both populations are stable under gene flow, because fit recombinant genotypes cannot be recreated. This is because ancestral alleles have been lost from both populations, or species, in the process of multi-step co-evolution (24,54). Another relevant aspect of positive within-population epistasis is that it becomes more likely as speciation progresses and more differences accumulate, analogous to the 'snowball' effect, but for positive interactions. This increases the chances that mutations whose direct effects are deleterious can fix if their positive epistatic or pleiotropic effects outweigh their direct negative effects. Such mutations would cause strongly negative fitness consequences when these positive epistatic effects are broken in hybrids. As a result, within-population epistasis could explain a snowball-like effect that occurs not due to an acceleration in the *number* of new incompatibilities, but due to increasing *strength* of negative fitness consequences with time (53).

*1.2.4 The mechanistic basis of interactions can be used to determine the shape of the fitness landscape*

The potential for system drift in an interaction network (e.g. regulatory network) depends on the number of genotypes via which the network can produce phenotypes of equivalent fitness, and how accessible these genotypes are through mutations. This information is equivalent to the shape of the fitness landscape. Models that capture the biophysics of transcription factor-binding (10,39,56,57) implicitly define a fitness landscape, with an important property: there are inherently many more ways to produce a moderately fit binding site than an optimally fit one, implying large regions of genotype space of moderate fitness. There will therefore be more potential for system drift in populations with sub-optimal regulatory pathways, such as small populations in which purifying selection is less efficient. This highlights the value of models that consider the genotype-phenotype map: earlier models using a fixed probability that a pair of substitutions will be incompatible found no effect of population size when speciation is driven by drift alone (20). A perhaps counterintuitive result is that in large populations it is phenotypes under the weakest selection that are most likely to accumulate the incompatibilities, as those are the ones where genetic drift can still lead to the system drift (10). However, it remains to be studied how general such predictions will be under different assumptions about the strength of selection and number of loci contributing to the trait, and their precise interactions (note the criticisms of low-dimensional fitness landscapes in 1.1 above).

*Summary of PART 1*

Theoretical approaches to study the accumulation of RI vary in their level of mechanistic detail, generality and tractability. From the existing theoretical work, some key findings on the role of multi-locus interactions in the build-up of RI emerge. Broadly speaking, models with and without a mechanistic understanding of multi-locus interactions both show that redundancy in the genotype (i.e. the fact that the same phenotype can be produced by multiple combinations of genotypes) allows incompatibilities to accumulate. In both contexts, predictions about how easily different genotypes can be reached through neutral evolution, and therefore how rapidly strong RI can evolve, are determined by the shape of the fitness landscape. Taking into account the mechanistic understanding and structure of multi-locus interactions allows for some additional insights. First, the mechanistic basis of interactions could itself predict the shape of the fitness landscape, and therefore affect the likelihood of incompatibility accumulation. Future work should aim to understand how features of realistic genotype-phenotype maps relate to features of abstract fitness landscapes (58), and incorporate the empirical advances in understanding fitness landscapes (59,60) into the theoretical models of speciation. Second, the number of interactions between loci influence the rate of accumulation of incompatibilities, and heterogeneity in interactions within a network (*i.e.* some nodes are highly connected, whereas others are not) could result in high variance in the rate of accumulation of incompatibilities and in speciation probability. This should be explored in further theoretical work. Third, insights from the mechanistic models could allow us to predict not only which kinds of loci are most likely to harbour incompatibilities (e.g. highly connected, central nodes or nodes that connect modules) but also whether the incompatibilities are likely to persist in the face of gene flow. Models that investigate the persistence of incompatibilities in the face of ongoing gene flow are necessary to reveal the long-term consequences of incompatibilities for RI, and should therefore be a focus of future work. Furthermore, in light of this special issue (see Introductory article), future work should compare scenarios where RI is weak to those of strong RI, and study whether the role of different types of incompatibilities in increasing RI differs between them. Finally, to be able to determine which models are most compatible with natural systems we need tools to connect predictions from theories to empirical patterns seen in genome-wide data. This could be achieved with simulations (54,61), which are discussed in more detail in Part 3. Next, we will discuss how multi-locus interactions can be examined in empirical speciation studies.

**PART 2: How can multi-locus interactions be investigated in experimental speciation studies?**

Mapping of incompatibility loci using classical genetic techniques in model organisms have revealed how epistatic fitness effects can produce strong reproductive barriers. For example, *Nup160* in *Drosophila* interacts with one or more unknown additional factors in the autosomal background (62), and *DM2* that underlies hybrid necrosis in *Arabidopsis* interacts with at least five different loci causing necrosis and problems in hybrids (63,64). In a few cases, these mapping approaches have been extended to whole chromosome or whole genome scale (65,66), revealing that there may be thousands of incompatibilities, many of which may have small effects, that contribute to species barriers. However, such studies are limited not only in their resolution but also to species where elaborate crossing experiments are possible. Recent developments in our ability to acquire genome-wide genetic or expression data for population samples of non-model organisms now provide the potential to extend this field to detect signatures of multi-locus interactions in non-model systems. In this section, we discuss genomic studies of naturally admixed populations as well as hybrid gene misexpression studies that reveal putatively disrupted protein-protein or regulatory interactions. We describe what such studies are beginning to reveal about the nature of species barriers, and we also highlight challenges in using these approaches.

*2.1 Admixed populations represent natural experiments where epistasis could be detected*

Hybrid zones and admixed populations in which fit and unfit combinations of alleles continue to segregate in a single population provide enhanced power and resolution to identify interactions that shape hybrid fitness. Turner and Harr (67) performed a genome-wide association study (GWAS) for traits associated with sterility in the house mouse hybrid zone. They found that most sterility-associated loci interact with more than one partner locus, and suggested that the variation in effect size among loci is correlated with the number of different networks in which the gene participates. These findings imply that models of speciation in which all pairs of loci can potentially interact (with a fixed probability of producing an incompatibility) (22,23) are inaccurate, and models in which only certain interactions are possible (46) are more likely to be realistic, due to the actual structure of the interactions.

When traits under selection in hybrids are not known, it is still possible to exploit admixed populations using naive scans for the effects of epistatic selection. Ancestry Disequilibrium ('AD') scans attempt to identify pairs of loci at which there are excessive statistical associations between allelic ancestry in admixed populations (68,69). In principle, AD scans could identify pairs of interacting loci that lead to fitness breakdown in hybrids and thereby reveal the architecture of barriers to gene flow and, together with genome annotation, the possible gene networks that underlie the barriers. Due to the requirement of fertile hybrids and enough divergence to cause negative epistasis, AD scans are likely to be most useful mid to later rather than early on in the speciation process. However, within-species detection of weak epistatic incompatibilities can be achieved with sufficient power (68), and may be particularly sensitive when candidate loci are known, such as mitochondria and their nuclear interacting partners (70). Using an AD scan, Pool (68) found evidence for epistasis and hub-like (one-to-many) interactions causing reduced fitness in an admixed *Drosophila melanogaster* population. One shortfall of AD scans is that they require existing polymorphism at both loci in the admixed population in order to detect ancestry disequilibrium. Therefore, the strongest incompatibilities may go undetected as selection could already have led to the fixation of one parental genotype at both loci, leaving only the weaker incompatibilities to be detected (61,68,71). Other concerns are that multiple testing of all possible pairs of loci leads to a high likelihood of false positives just due to chance, and that the AD scans assume random mating within the population, as non-random mating will also result in statistical association between unlinked regions of the genome in mixed populations. These issues can be partly alleviated by examining the overlap between candidate incompatibilities in independent admixed populations (69), identification of 'hub-like' interactions that involve the same gene multiple times; or investigating whether the candidate genes are known to participate in the same pathways (68). For example, a recent study documented signatures of epistatic selection on archaic introgression in humans by combining information on introgressed genes, their co-segregation and functional information about the biological pathways in which they function (72).

*2.2 Genome scans combined with additional functional information could reveal multi-locus interactions*

Even when natural hybrids are rare or absent, analysis of genetic differentiation along the genome between diverging populations can reveal loci at which selection has resisted genetic exchange between the populations in the past, therefore indicating likely loci contributing to the genomic barrier (73–76). Genome scans are easy to perform, and in contrast to AD scans do not require that pairs of incompatible alleles are segregating as polymorphisms in a hybrid population. Furthermore, genome scans are uniquely able to detect signals of weak selection accumulated over multiple generations which could be missed by experimental studies or AD scans that focus on selection over a single generation (77). However, genome scans become less useful the further the speciation process proceeds as the signatures of selection against foreign alleles may be difficult to distinguish from those of confounding processes such as purifying selection acting within the diverging lineages (78–80). Estimating effective migration rate instead of divergence along the genome holds promise to increase our understanding of later stages of the speciation process in the future (18,76,77). Unlike incompatibilities that are also involved in local adaptation, those incompatibilities whose fitness effects depend only on the genetic background will not necessarily produce localised barriers that persist in the face of gene flow, particularly if it is possible to 'rescue' hybrid fitness through recombination to regenerate fit combinations of alleles (24,54,81). By contrast, incompatibilities that emerge through multi-step co-evolution between interacting loci in both populations can generate persistent, localised barriers to gene flow (54). Therefore, this form of co-evolved incompatibility is most likely to be detectable by genome scans.

Unlike AD scans, differentiation scans do not directly reveal whether epistasis between loci underlies barriers to gene flow. Instead, the genome scan can be used as a first step in inference, after which other methods are needed to test for interactions at the molecular level, or epistasis. One approach to make these links is to test whether candidate barrier loci show experimental evidence for epistasis in natural or artificial hybrids. Such an epistatic interaction has been demonstrated between loci underlying plumage divergence in crows (82). Another approach is to ask whether the candidate loci show interactions at the molecular level, which could be a potential indicator of epistatic interactions, an approach utilized by Kulmuni et al. (83). Using information from the String database (84) they found that a large proportion of putative barrier loci identified between hybridizing species of wood ants are known to form part of a single interaction network in *Drosophila*. Interestingly, random samples from the same marker set (including loci not thought to be involved in barriers), identified similar interactivity. This might imply that any large-enough gene set would include many members known to interact biologically, hampering the identification of true epistatic interactions underlying RI. There are also empirical methods to detect protein-protein interactions, like pull-down assays (reviewed in (85)), that can be utilized after identification of candidate genes. However, we still know very little about how interactions between loci at the molecular level translate into epistasis that is relevant for specific traits or individual fitness.

*2.3 Gene expression can be utilized to find disrupted gene regulation in hybrids*

Phenotypic evolution and adaptation involve divergence in gene regulation. Particularly early in divergence, regulatory differences appear to accumulate more rapidly than amino acid substitutions in protein sequences (86,87). Gene expression is controlled by cis- and trans-acting factors, and co-evolution between regulatory factors within a lineage can result in misexpression and breakdown in hybrids (55,88). These findings are consistent with models described in Part 1.2, in which regulatory changes can emerge at multiple loci without a change in the overall phenotype. The combination of such diverged regulatory networks in hybrids can cause their gene expression profiles to be distinct from those of both parental populations. Several recent studies have described gene 'mis-regulation' in hybrids (e.g. (89–93)). Mis-regulation in F1 hybrids is indicative of epistasis that could underlie reduced hybrid fitness, and therefore RI. Although hundreds or even thousands of individual genes can be misexpressed in hybrids, these differences could come about through modifications at a smaller number of key regulatory nodes in complex networks (94). This is shown by Turner et al. (93) who mapped interactions between expression QTLs and genotypes in the case of house mouse hybrid sterility. They found complex regulatory interactions across the genome, with a single eQTL interacting with 17 to >1000 partners, suggesting that regulatory divergence at many genes could be explained by evolution of a relatively small number of master regulators. Gene expression data can also be used to reveal networks of co-expressed genes (95). Comparing co-expression networks in hybrids and parental populations can reveal whole networks disrupted in hybrids as opposed to individual genes (96). While gene expression studies in hybrids reveal possible candidates for gene interactions underlying speciation, they do not directly demonstrate the fitness consequences of changes in hybrid expression. Indeed, hybrids are sometimes more fit than the parental populations. Making the links between regulatory interactions, hybrid misexpression and fitness can be achieved by comparing multi-locus genotypes and their gene expression patterns in fit and unfit classes of hybrids (93).

*Summary of PART 2*

Future studies that combine genome scan, AD scan and gene expression analyses in hybrids are likely to help in detecting interactions underlying RI. Using any one of these approaches alone can be biased by false positives or confounding signals, but overlapping information from multiple approaches can help to narrow down candidate interactions. Genome scans will be most useful early on in speciation, AD scans later on in the process (provided admixed populations are available) and hybrid misexpression and co-expression network studies potentially informative at any point in the speciation continuum. Where possible, signatures of selection against incompatible combinations of alleles could be measured in real-time, e.g. by comparing early and late developmental stages, which could provide further evidence for interactions (97). On top of that, if the time window of hybrid breakdown is known, gene co-expression network analyses could identify networks disrupted in hybrids. A completely opposite approach to these different types of scans would be to investigate classes of loci with high or low levels of known physical interactions and characterize average levels of differentiation and inferred rates of effective gene flow for these. As Shih (47) demonstrated with undirected (e.g. PPI) networks, this data can reveal whether physically highly connected genes are more likely to be involved in interactions disrupted by mutations than less connected genes on average. Finally, in the future there is a clear need for approaches connecting mechanistic understanding of gene interactions, population genetic theory and empirical genomic data.

**PART 3: Conclusions and future directions**

The idea that gene interactions and epistasis are central to speciation is over a century old. Much of the progress has been made by modelling epistasis without a detailed mechanistic understanding of gene interactions. Recent models that explicitly consider the structure of multi-locus interactions and how they produce phenotypes, have in many ways recapitulated earlier results, but have allowed prediction of some novel patterns that are consistent with empirical data. These include a high variance in the rate of build-up of RI among taxa, the emergence of strong incompatibilities that produce localised barriers to introgression, and an effect of population size on the build-up of RI. Experimental approaches that use genome-scale data have proved useful to detect putative interactions leading to epistatic fitness effects and RI, but they have generally been utilized in isolation. Combining ancestry disequilibrium (AD) scans, genome scans, hybrid gene expression and assays of hybrid phenotypes and fitness could overcome the shortfalls of individual methods in the future, and provide stronger evidence for the involvement of particular loci in interactions resulting in hybrid breakdown. In addition to identifying particular interactions, ultimately, one would like to test predictions from theory using empirical genomic data. In order to do this, we need to predict what kind of patterns incompatibilities will leave in genome-wide polymorphism data, if any. Thus, we propose a key need for the future: studies and tools that link between theory and experimental studies. Cross-talk between theory and empirical work can facilitate interpreting genome scan results, for example, but also help point to gene characteristics (like pleiotropy) that should be enriched among barrier loci.

Future studies should help bridge theory and empirical work by answering the following needs. First, there is a need to test the persistence of multi-locus incompatibilities in the face of long-term gene flow, which is likely in nature and will determine what patterns incompatibilities could leave in genomic data (54). An additional aim here is to determine whether genomic approaches would have sufficient power to detect multi-locus incompatibilities (61). Second, there is a need for testable predictions about the patterns different processes of incompatibility accumulation leave at the genomic level. This is important as models incorporating different selective and genetic mechanisms (snowball models, system drift models, tipping point models (98) etc.) may predict similar broad-scale patterns, such as faster than linear accumulation of incompatibilities at some point of the speciation process. However, different processes of incompatibility accumulation may leave different signatures in the genome. For example, neutral or nearly neutral processes are unlikely to cause large selective sweeps in genomes, as have been seen under local adaptation based on few major effect loci (99). Third, there is a need to model highly-polygenic and pleiotropic genome-wide architecture of traits (*i.e.* the omnigenic model (45)) leading to incompatibilities. This is relevant, because the effects of such a complex polygenic architecture will be distinct from the effects of architectures with fewer loci. Indeed, even polygenic models without interactions can reveal non-linearities not observed in models with small numbers of loci (98). Finally, there is a need to model the temporal time scale to understand the processes occurring early versus later in the build-up of RI, as the relative role of multi-locus interactions could change as speciation progresses.

One potential tool to bridge between theory and observed genomic data are simulations. This is now achievable thanks to the availability of efficient tools for both coalescent (100) and forward genetic (101) simulations in a genome-wide setting with large numbers of selected loci and complex selection regimes (101). Simulations could begin to answer how incompatibilities arising from multi-locus interactions translate into patterns that could be detected in population genomic studies, an approach that was successfully utilized e.g. in (54). Furthermore, studies not focusing on epistasis per se have successfully employed simulations to investigate RI in a genome- or chromosome-wide setting (16,102–104). Importantly, simulations also allow us to account for the noisy nature of genomic data: each speciation event represents a single realisation of a stochastic evolutionary process (105). Simulations could be utilized in conjunction with empirical data to fit different models of incompatibility accumulation, gene flow and demography to make inferences about the nature of real species barriers, and importantly ask if the observed data could be explained by a neutral model. They could potentially allow distinguishing diagnostic differences between multi-locus incompatibilities and confounding processes.

Looking beyond speciation, there is also a need to better understand multi-locus interactions. How do protein interaction networks relate to regulatory networks and how do these translate into fitness landscapes? How do multi-locus interactions evolve over short and long evolutionary time scales and how do changes in network topology and connectivity affect phenotypes and fitness? Much progress in this direction has been made recently (e.g. (7,8,106–108)).

**Acknowledgments**

We would like to thank Dorothea Lindtke, Tobias Uller, Pierre Nouhaud, two anonymous reviewers and the editor Anja Westram for useful feedback that greatly improved our manuscript.